# 1. BIG-BANG NUCLEOSYNTHESIS

Revised October 2005 by B.D. Fields (Univ. of Illinois) and S. Sarkar (Univ. of Oxford).

Big-bang nucleosynthesis (BBN) offers the deepest reliable probe of the early universe, being based on well-understood Standard Model physics [1–4]. Predictions of the abundances of the light elements, D, $^3$He, $^4$He, and $^7$Li, synthesized at the end of the "first three minutes" are in good overall agreement with the primordial abundances inferred from observational data, thus validating the standard hot big-bang cosmology (see [5] for a review). This is particularly impressive given that these abundances span nine orders of magnitude — from $^4$He/H $\sim 0.08$ down to $^7$Li/H $\sim 10^{-10}$ (ratios by number). Thus BBN provides powerful constraints on possible deviations from the standard cosmology [2], and on new physics beyond the Standard Model [3].

## 1.1. Theory

The synthesis of the light elements is sensitive to physical conditions in the early radiation-dominated era at temperatures $T \lesssim 1$ MeV, corresponding to an age $t \gtrsim 1$ s. At higher temperatures, weak interactions were in thermal equilibrium, thus fixing the ratio of the neutron and proton number densities to be $n/p = e^{-Q/T}$, where $Q = 1.293$ MeV is the neutron-proton mass difference. As the temperature dropped, the neutron-proton inter-conversion rate, $\Gamma_{n\leftrightarrow p} \sim G_F^2 T^5$, fell faster than the Hubble expansion rate, $H \sim \sqrt{g_* G_N}\, T^2$, where $g_*$ counts the number of relativistic particle species determining the energy density in radiation. This resulted in departure from chemical equilibrium ("freeze-out") at $T_{\rm fr} \sim (g_* G_N/G_F^4)^{1/6} \simeq 1$ MeV. The neutron fraction at this time, $n/p = e^{-Q/T_{\rm fr}} \simeq 1/6$, is thus sensitive to every known physical interaction, since $Q$ is determined by both strong and electromagnetic interactions while $T_{\rm fr}$ depends on the weak as well as gravitational interactions. Moreover the sensitivity to the Hubble expansion rate affords a probe of e.g. the number of relativistic neutrino species [6]. After freeze-out the neutrons were free to $\beta$-decay so the neutron fraction dropped to $\simeq 1/7$ by the time nuclear reactions began. A simplified analytic model of freeze-out yields the $n/p$ ratio to an acuracy of $\sim 1\%$ [7,8].

The rates of these reactions depend on the density of baryons (strictly speaking, nucleons), which is usually expressed normalized to the relic blackbody photon density as $\eta \equiv n_B/n_\gamma$. As we shall see, all the light-element abundances can be explained with $\eta_{10} \equiv \eta \times 10^{10}$ in the range 4.7–6.5 (95% CL). With $n_\gamma$ fixed by the present CMB temperature 2.725 K (see Cosmic Microwave Background review), this can be stated as the allowed range for the baryon mass density today, $\rho_B = (3.2\text{--}4.5) \times 10^{-31}$ g cm$^{-3}$, or as the baryonic fraction of the critical density, $\Omega_B = \rho_B/\rho_{\rm crit} \simeq \eta_{10} h^{-2}/274 = (0.017\text{--}0.024) h^{-2}$, where $h \equiv H_0/100$ km s$^{-1}$ Mpc$^{-1}$ = $0.72 \pm 0.08$ is the present Hubble parameter (see Cosmological Parameters review).

The nucleosynthesis chain begins with the formation of deuterium in the process $p(n,\gamma)$D. However, photo-dissociation by the high number density of photons delays production of deuterium (and other complex nuclei) well after $T$ drops below the binding energy of deuterium, $\Delta_D = 2.23$ MeV. The quantity $\eta^{-1} e^{-\Delta_D/T}$, i.e. the number of photons per baryon above the deuterium photo-dissociation threshold, falls below unity at $T \simeq 0.1$ MeV; nuclei can then begin to form without being immediately photo-dissociated again. Only 2-body reactions such as D($p, \gamma$)$^3$He, $^3$He(D, $p$)$^4$He, are important because the density has become rather low by this time.

Nearly all the surviving neutrons when nucleosynthesis begins end up bound in the most stable light element $^4$He. Heavier nuclei do not form in any significant quantity both because of the absence of stable nuclei with mass number 5 or 8 (which impedes nucleosynthesis via $n^4$He, $p^4$He or $^4$He$^4$He reactions) and the large Coulomb barriers for reactions such as T($^4$He, $\gamma$)$^7$Li and $^3$He($^4$He, $\gamma$)$^7$Be. Hence the primordial mass fraction of $^4$He, conventionally referred to as $Y_\mathrm{p}$, can be estimated by the simple counting argument

$$Y_\mathrm{p} = \frac{2\,(n/p)}{1 + n/p} \simeq 0.25 \;. \tag{1.1}$$

There is little sensitivity here to the actual nuclear reaction rates, which are however important in determining the other "left-over" abundances: D and $^3$He at the level of a few times $10^{-5}$ by number relative to H, and $^7$Li/H at the level of about $10^{-10}$ (when $\eta_{10}$ is in the range 1–10). These values can be understood in terms of approximate analytic arguments [8,9]. The experimental parameter most important in determining $Y_\mathrm{p}$ is the neutron lifetime, $\tau_n$, which normalizes (the inverse of) $\Gamma_{n \not\to p}$. (This is not fully determined by $G_\mathrm{F}$ alone since neutrons and protons also have strong interactions, the effects of which cannot be calculated very precisely.) The experimental uncertainty in $\tau_n$ used to be a source of concern but has recently been reduced substantially: $\tau_n = 885.7 \pm 0.8$ s (see N Baryons listing).

The elemental abundances, calculated using the (publicly available [10]) Wagoner code [1,11], are shown in Fig. 1.1 as a function of $\eta_{10}$. The $^4$He curve includes small corrections due to radiative processes at zero and finite temperature [12], non-equilibrium neutrino heating during $e^\pm$ annihilation [13], and finite nucleon mass effects [14]; the range reflects primarily the $1\sigma$ uncertainty in the neutron lifetime. The spread in the curves for D, $^3$He and $^7$Li corresponds to the $1\sigma$ uncertainties in nuclear cross sections estimated by Monte Carlo methods [15–16]. Recently the input nuclear data have been carefully reassessed [17–21], leading to improved precision in the abundance predictions. Polynomial fits to the predicted abundances and the error correlation matrix have been given [16,22]. The boxes in Fig. 1.1 show the observationally inferred primordial abundances with their associated statistical and systematic uncertainties, as discussed below.

## 1.2. Light Element Abundances

BBN theory predicts the universal abundances of D, $^3$He, $^4$He, and $^7$Li, which are essentially determined by $t \sim 180$ s. Abundances are however observed at much later epochs, after stellar nucleosynthesis has commenced. The ejected remains of this stellar processing can alter the light element abundances from their primordial values, but also produce heavy elements such as C, N, O, and Fe ("metals"). Thus one seeks astrophysical sites with low metal abundances, in order to measure light element abundances which are closer to primordial. For all of the light elements, systematic errors are an important and often dominant limitation to the precision with which primordial abundances can be inferred.

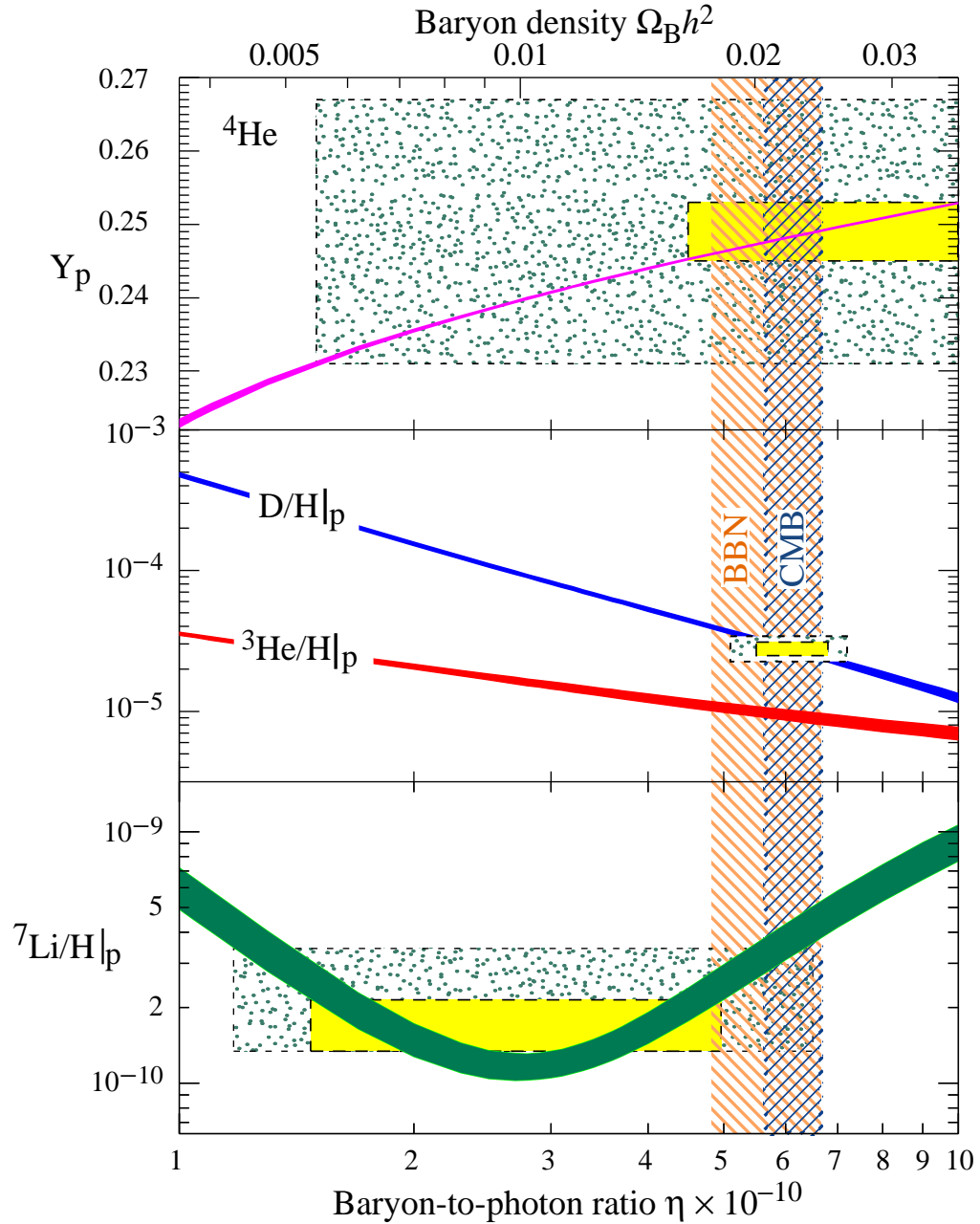

**Figure 1.1:** The abundances of $^4$He, D, $^3$He and $^7$Li as predicted by the standard model of big-bang nucleosynthesis. Boxes indicate the observed light element abundances (smaller boxes: $2\sigma$ statistical errors; larger boxes: $\pm 2\sigma$ statistical and systematic errors). The narrow vertical band indicates the CMB measure of the cosmic baryon density. See full-color version on color pages at end of book.

In recent years, high-resolution spectra have revealed the presence of D in high-redshift, low-metallicity quasar absorption systems (QAS), via its isotope-shifted Lyman-$\alpha$ absorption [23–27]. It is believed that there are no astrophysical sources of deuterium [28], so any detection provides a lower limit to primordial D/H and thus an upper limit on $\eta$; for example, the local interstellar value of D/H $= (1.5 \pm 0.1) \times 10^{-5}$ [29] requires $\eta_{10} \leq 9$. Recent observations find an unexpected scatter of a factor of $\sim 2$ [30] so interstellar D may have been depleted by stellar processing. However, for the high-redshift systems, conventional models of galactic nucleosynthesis (chemical evolution) do not predict D/H depletion [31].

The observed extragalactic D values are bracketed by the non-detection of D in a high-redshift system, D/H $< 6.7 \times 10^{-5}$ at $1\sigma$ [32], and low values in some (damped Lyman-$\alpha$) systems [24,25]. Averaging the 5 most precise observations of deuterium in QAS gives D/H $= (2.78 \pm 0.15) \times 10^{-5}$, where the error is statistical only [23,26]. However there remains concern over systematic errors, the dispersion between the values being much larger than is expected from the individual measurement errors ($\chi^2 = 16.5$ for $\nu = 4$ d.o.f.). Increasing the error by a factor $\sqrt{\chi^2/\nu}$ gives, as shown on Fig. 1.1:

$$\text{D/H} = (2.78 \pm 0.29) \times 10^{-5}. \tag{1.2}$$

We observe $^4$He in clouds of ionized hydrogen (H II regions), the most metal-poor of which are in dwarf galaxies. There is now a large body of data on $^4$He and CNO in these systems [33]. These data confirm that the small stellar contribution to helium is positively correlated with metal production. Extrapolating to zero metallicity gives the primordial $^4$He abundance [34]

$$Y_{\text{p}} = 0.249 \pm 0.009. \tag{1.3}$$

Here the latter error is a careful (and significantly enlarged) estimate of the systematic uncertainties; these dominate, and is based on the scatter in different analyses of the physical properties of the H II regions [33,34]. Other extrapolations to zero metallicity give $Y_{\text{p}} = 0.2443 \pm 0.0015$ [33], and $Y_{\text{p}} = 0.2391 \pm 0.0020$ [35]. These are consistent (given the systematic errors) with the above estimate [34], which appears in Fig. 1.1.

The systems best suited for Li observations are metal-poor stars in the spheroid (Pop II) of our Galaxy, which have metallicities going down to at least $10^{-4}$ and perhaps $10^{-5}$ of the Solar value [36]. Observations have long shown [37–41] that Li does not vary significantly in Pop II stars with metallicities $\lesssim 1/30$ of Solar — the "Spite plateau" [37]. Recent precision data suggest a small but significant correlation between Li and Fe [38] which can be understood as the result of Li production from Galactic cosmic rays [39]. Extrapolating to zero metallicity one arrives at a primordial value Li/H$|_{\text{p}} = (1.23 \pm 0.06) \times 10^{-10}$ [40]. One systematic error stems from the differences in techniques to determine the physical parameters (e.g., the temperature) of the stellar atmosphere in which the Li absorption line is formed. Alternative analyses, using methods that give systematically higher temperatures, yield Li/H$|_{\text{p}} = (2.19 \pm 0.28) \times 10^{-10}$ [41] and Li/H$|_{\text{p}} = (2.34 \pm 0.32) \times 10^{-10}$ [42]; the difference with [40] indicates a systematic uncertainty of a factor of $\sim 2$. Moreover it is possible that the Li in Pop II stars has been partially destroyed, due to mixing of the outer layers with the hotter interior [43]. Such processes can be constrained by the absence of significant scatter in Li-Fe [38], and by observations of the fragile isotope $^6$Li [39]. Nevertheless, depletion by a factor as large as $\sim 1.6$ remains allowed [40,43]. Including these systematics, we estimate a primordial Li range, as shown in Fig. 1.1:

$$\text{Li/H}|_\text{p} = \left(1.7 \pm 0.02^{+1.1}_{-0}\right) \times 10^{-10}. \tag{1.4}$$

Finally, we turn to $^3$He. Here, the only observations available are in the Solar system and (high-metallicity) H II regions in our Galaxy [44]. This makes inference of the primordial abundance difficult, a problem compounded by the fact that stellar nucleosynthesis models for $^3$He are in conflict with observations [45]. Consequently, it is no longer appropriate to use $^3$He as a cosmological probe; instead, one might hope to turn the problem around and constrain stellar astrophysics using the predicted primordial $^3$He abundance [46].

## 1.3. Concordance, Dark Matter, and the CMB

We now use the observed light element abundances to assess the theory. We first consider standard BBN, which is based on Standard Model physics alone, so $N_\nu = 3$ and the only free parameter is the baryon-to-photon ratio $\eta$. (The implications of BBN for physics beyond the Standard Model will be considered below, §4). Thus, any abundance measurement determines $\eta$, while additional measurements overconstrain the theory and thereby provide a consistency check.

First we note that the overlap in the $\eta$ ranges spanned by the larger boxes in Fig. 1.1 indicates overall concordance. More quantitatively, when we account for theoretical uncertainties as well as the statistical and systematic errors in observations, there is acceptable agreement among the abundances when

$$4.7 \leq \eta \leq 6.5 \ (95\% \ \text{CL}). \tag{1.5}$$

However the agreement is much less satisfactory if we use only the quoted statistical errors in the observations. In particular, as seen in Fig. 1.1, D and $^4$He are consistent with each other but favour a value of $\eta$ which is higher by $\sim 2\sigma$ from that indicated by the $^7$Li abundance. Additional studies are required to clarify if this discrepancy is real.

Even so, the overall concordance is remarkable: using well-established microphysics we have extrapolated back to an age of $\sim 1$ s to correctly predict light element abundances spanning 9 orders of magnitude. This is a major success for the standard cosmology, and inspires confidence in extrapolation back to still earlier times.

This concordance provides a measure of the baryon content

$$0.017 \leq \Omega_\text{B} h^2 \leq 0.024 \ (95\% \ \text{CL}) \ , \tag{1.6}$$

a result that plays a key role in our understanding of the matter budget of the universe. First we note that $\Omega_B \ll 1$, i.e., baryons cannot close the universe [47]. Furthermore, the cosmic density of (optically) luminous matter is $\Omega_{\text{lum}} \simeq 0.0024h^{-1}$ [48], so that $\Omega_B \gg \Omega_{\text{lum}}$: most baryons are optically dark, probably in the form of a $\sim 10^6$ K X-ray emitting intergalactic medium [49]. Finally, given that $\Omega_M \sim 0.3$ (see Dark Matter and Cosmological Parameters reviews), we infer that most matter in the universe is not only dark but also takes some non-baryonic (more precisely, non-nucleonic) form.

The BBN prediction for the cosmic baryon density can be tested through precision observations of CMB temperature fluctuations (see Cosmic Microwave Background review). One can determine $\eta$ from the amplitudes of the acoustic peaks in the CMB angular power spectrum [50], making it possible to compare two measures of $\eta$ using very different physics, at two widely separated epochs. In the standard cosmology, there is no change in $\eta$ between BBN and CMB decoupling, thus, a comparison of $\eta_{\text{BBN}}$ and $\eta_{\text{CMB}}$ is a key test. Agreement would endorse the standard picture while disagreement could point to new physics during/between the BBN and CMB epochs.

The release of the first-year WMAP results was a landmark event in this test of BBN. As with other cosmological parameter determinations from CMB data, the derived $\eta_{\text{CMB}}$ depends on the adopted priors [51], in particular the form assumed for the power spectrum of primordial density fluctuations. If this is taken to be a scale-free power-law, the WMAP data implies $\Omega_B h^2 = 0.024 \pm 0.001$ or $\eta_{10} = 6.58 \pm 0.27$, while allowing for a "running" spectral index lowers the value to $\Omega_B h^2 = 0.0224 \pm 0.0009$ or $\eta_{10} = 6.14 \pm 0.25$ [52]; this latter range appears in Fig. 1.1. Other assumptions for the shape of the power spectrum can lead to baryon densities as low as $\Omega_B h^2 = 0.019$ [53]. Thus outstanding uncertainties regarding priors are a source of systematic error which presently exceeds the statistical error in the prediction for $\eta$.

It is remarkable that the CMB estimate of the baryon density is consistent with the BBN range quoted in Eq. (1.6), and is in fact in very good agreement with the value inferred from recent high-redshift D/H measurements [26] and $^4$He determinations; together these observations span diverse envirnoments from redshifts $z = 1000$ to the present. However $^7$Li is at best marginally consistent with the CMB (and with D) given the error budgets we have quoted. The question then becomes more pressing as to whether this mismatch come from systematic errors in the observed abundances, and/or uncertainties in stellar astrophysics, or whether there might be new physics at work. Inhomogeneous nucleosynthesis can alter abundances for a given $\eta_{\text{BBN}}$ but will overproduce $^7$Li [54]. Note that entropy generation by some non-standard process could have decreased $\eta$ between the BBN era and CMB decoupling, however the lack of spectral distortions in the CMB rules out any significant energy injection upto a redshift $z \sim 10^7$ [55]. Interestingly, the CMB itself offers the promise of measuring $^4$He [56] and possibly $^7$Li [57] directly at $z \sim 300 - 1000$.

Bearing in mind the importance of priors, the promise of precision determinations of the baryon density using the CMB motivates using this value as an input to BBN calculations. Within the context of the Standard Model, BBN then becomes a zero-parameter theory, and the light element abundances are completely determined to within the uncertainties in $\eta_{\text{CMB}}$ and the BBN theoretical errors. Comparison with the observed abundances then can be used to test the astrophysics of post-BBN light element evolution [58]. Alternatively, one can consider possible physics beyond the Standard Model (e.g., which might change the expansion rate during BBN) and then use all of the abundances to test such models; this is the subject of our final section.

## 1.4. Beyond the Standard Model

Given the simple physics underlying BBN, it is remarkable that it still provides the most effective test for the cosmological viability of ideas concerning physics beyond the Standard Model. Although baryogenesis and inflation must have occurred at higher temperatures in the early universe, we do not as yet have 'standard models' for these so BBN still marks the boundary between the established and the speculative in big bang cosmology. It might appear possible to push the boundary back to the quark-hadron transition at $T \sim \Lambda_{\rm QCD}$ or electroweak symmetry breaking at $T \sim 1/\sqrt{G_{\rm F}}$; however so far no observable relics of these epochs have been identified, either theoretically or observationally. Thus although the Standard Model provides a precise description of physics up to the Fermi scale, cosmology cannot be traced in detail before the BBN era.

Limits on particle physics beyond the Standard Model come mainly from the observational bounds on the $^4$He abundance. This is proportional to the $n/p$ ratio which is determined when the weak-interaction rates fall behind the Hubble expansion rate at $T_{\rm fr} \sim 1$ MeV. The presence of additional neutrino flavors (or of any other relativistic species) at this time increases $g_*$, hence the expansion rate, leading to a larger value of $T_{\rm fr}$, $n/p$, and therefore $Y_{\rm p}$ [6,59]. In the Standard Model, the number of relativistic particle species at 1 MeV is $g_* = 5.5 + \frac{7}{4} N_\nu$, where 5.5 accounts for photons and $e^\pm$, and $N_\nu$ is the number of (nearly massless) neutrino flavors (see Big Bang Cosmology review). The helium curves in Fig. 1.1 were computed taking $N_\nu = 3$; the computed abundance scales as $\Delta Y_{\rm BBN} \simeq 0.013 \Delta N_\nu$ [7]. Clearly the central value for $N_\nu$ from BBN will depend on $\eta$, which is independently determined (with weaker sensitivity to $N_\nu$) by the adopted D or $^7$Li abundance. For example, if the best value for the observed primordial $^4$He abundance is 0.249, then, for $\eta_{10} \sim 6$, the central value for $N_\nu$ is very close to 3. This limit depends sensitively on the adopted light element abudances, particularly $Y_{\rm BBN}$. A maximum likelihood analysis on $\eta$ and $N_\nu$ based on the above $^4$He and D abundances finds the (correlated) 95% CL ranges to be $4.9 < \eta_{10} < 7.1$ and $1.8 < N_\nu < 4.5$ [60]. On the other hand, using the 'low' $^4$He and $^7$Li gives [61] $\eta_{10} \leq 4.3$, and $1.4 \leq N_\nu \leq 4.9$. Similar results were obtained in another study [62] which presented a simpler method [10] to extract such bounds based on $\chi^2$ statistics, given a set of input abundances. Using the CMB determination of $\eta$ improves the constraints: with a 'low' $^4$He, $N_\nu = 3$ is barely allowed at $2\sigma$ [63] but using the $^4$He (and D) abundance quoted above gives $5.66 < \eta_{10} < 6.58$ ($\Omega_{\rm B} h^2 = 0.0226 \pm 0.0017$) and $N_\nu = 3.24 \pm 1.2$ at 95% CL [60].

Just as one can use the measured helium abundance to place limits on $g_*$ [59], any changes in the strong, weak, electromagnetic, or gravitational coupling constants, arising e.g. from the dynamics of new dimensions, can be similarly constrained [64].

The limits on $N_\nu$ can be translated into limits on other types of particles or particle masses that would affect the expansion rate of the Universe during nucleosynthesis.

For example consider 'sterile' neutrinos with only right-handed interactions of strength $G_R < G_F$. Such particles would decouple at higher temperature than (left-handed) neutrinos, so their number density ($\propto T^3$) relative to neutrinos would be reduced by any subsequent entropy release, e.g. due to annihilations of massive particles that become non-relativistic in between the two decoupling temperatures. Thus (relativistic) particles with less than full strength weak interactions contribute less to the energy density than particles that remain in equilibrium up to the time of nucleosynthesis [65]. If we impose $N_\nu < 4$ as an illustrative constraint, then the three right-handed neutrinos must have a temperature $3(T_{\nu_R}/T_{\nu_L})^4 < 1$. Since the temperature of the decoupled $\nu_R$'s is determined by entropy conservation (see Big Bang Cosmology review), $T_{\nu_R}/T_{\nu_L} = [(43/4)/g_*(T_d)]^{1/3} < 0.76$, where $T_d$ is the decoupling temperature of the $\nu_R$'s. This requires $g_*(T_d) > 24$ so decoupling must have occurred at $T_d > 140$ MeV. The decoupling temperature is related to $G_R$ through $(G_R/G_F)^2 \sim (T_d/3\,\mathrm{MeV})^{-3}$, where 3 MeV is the decoupling temperature for $\nu_L$s. This yields a limit $G_R \lesssim 10^{-2} G_F$. The above argument sets lower limits on the masses of new $Z'$ gauge bosons in superstring models [66] or in extended technicolour models [67] to which such right-handed neutrinos would be coupled. Similarly a Dirac magnetic moment for neutrinos, which would allow the right-handed states to be produced through scattering and thus increase $g_*$, can be significantly constrained [68], as can any new interactions for neutrinos which have a similar effect [69]. Right-handed states can be populated directly by helicity-flip scattering if the neutrino mass is large enough and this can be used to used to infer e.g. a bound of $m_{\nu_\tau} \lesssim 1$ MeV taking $N_\nu < 4$ [70]. If there is mixing between active and sterile neutrinos then the effect on BBN is more complicated [71].

The limit on the expansion rate during BBN can also be translated into bounds on the mass/lifetime of particles which are non-relativistic during BBN resulting in an even faster speed-up rate; the subsequent decays of such particles will typically also change the entropy, leading to further constraints [72]. Even more stringent constraints come from requiring that the synthesized light element abundances are not excessively altered through photodissociations by the electromagnetic cascades triggered by the decays [73–75], or by the effects of hadrons in the cascades [76]. Such arguments have been applied to e.g. rule out a MeV mass $\nu_\tau$ which decays during nucleosynthesis [77]; even if the decays are to non-interacting particles (and light neutrinos), bounds can be derived from considering their effects on BBN [78].

Such arguments have proved very effective in constraining supersymmetry. For example if the gravitino is light and contributes to $g_*$, the illustrative BBN limit $N_\nu < 4$ requires its mass to exceed $\sim 1$ eV [79]. In models where supersymmetry breaking is gravity mediated, the gravitino mass is usually much higher, of order the electroweak scale; such gravitinos would be unstable and decay after BBN. The constraints on unstable particles discussed above imply stringent bounds on the allowed abundance of such particles, which in turn impose powerful constraints on supersymmetric inflationary cosmology [74–76]. These can be evaded only if the gravitino is massive enough to decay before BBN, i.e. $m_{3/2} \gtrsim 50$ TeV [80] which would be unnatural, or if it is in fact the LSP and thus stable [74,81]. Similar constraints apply to moduli — very weakly coupled fields in string theory which obtain an electroweak-scale mass from supersymmetry breaking [82].

Finally we mention that BBN places powerful constraints on the recently suggested possibility that there are new large dimensions in nature, perhaps enabling the scale of quantum gravity to be as low as the electroweak scale [83]. Thus Standard Model fields may be localized on a 'brane' while gravity alone propagates in the 'bulk'. It has been further noted that the new dimensions may be non-compact, even infinite [84] and the cosmology of such models has attracted considerable attention. The expansion rate in the early universe can be significantly modified so BBN is able to set interesting constraints on such possibilities [85].

**References:**

1.  R.V. Wagoner *et al.*, Astrophys. J. **148**, 3 (1967).
2.  R.A. Malaney and G.J. Mathews, Phys. Reports **229**, 145 (1993).
3.  S. Sarkar, Rept. on Prog. in Phys. **59**, 1493 (1996).
4.  D.N. Schramm and M.S. Turner, Rev. Mod. Phys. **70**, 303 (1998).
5.  K.A. Olive *et al.*, Phys. Reports **333**, 389 (2000).
6.  P.J.E. Peebles, Phys. Rev. Lett. **16**, 411 (1966).
7.  J. Bernstein *et al.*, Rev. Mod. Phys. **61**, 25 (1989).
8.  S. Mukhanov, Int. J. Theor. Phys. **143**, 669 (2004).
9.  R. Esmailzadeh *et al.*, Astrophys. J. **378**, 504 (1991).
10. http://www-thphys.physics.ox.ac.uk/users/SubirSarkar/bbn.html.
11. L. Kawano, FERMILAB-PUB-92/04-A.
12. S. Esposito *et al.*, Nucl. Phys. **B568**, 421 (2000).
13. S. Dodelson and M.S. Turner, Phys. Rev. **D46**, 3372 (1992).
14. D. Seckel, hep-ph/9305311;
    R. Lopez and M.S. Turner, Phys. Rev. **D59**, 103502 (1999).
15. M.S. Smith *et al.*, Astrophys. J. Supp. **85**, 219 (1993).
16. G. Fiorentini *et al.*, Phys. Rev. **D58**, 063506 (1998).
17. K.M. Nollett and S. Burles, Phys. Rev. **D61**, 123505 (2000).
18. R.H. Cyburt *et al.*, New Astron. **6**, 215 (2001).
19. A. Coc *et al.*, Astrophys. J. **600**, 544 (2004).
20. R.H. Cyburt, Phys. Rev. **D70**, 023505 (2004).
21. P.D. Serpico *et al.*, J. Cosmo. Astropart. Phys. **12**, 010 (2004).
22. K.M. Nollett *et al.*, Astrophys. J. Lett. **552**, L1 (2001).
23. J.M. O'Meara *et al.*, Astrophys. J. **552**, 718 (2001).
24. S. D'Odorico *et al.*, Astron. & Astrophys. **368**, L21 (2001).
25. M. Pettini and D. Bowen, Astrophys. J. **560**, 41 (2001).
26. D. Kirkman *et al.*, Astrophys. J. Supp. **149**, 1 (2003).
27. N.H.M. Crighton *et al.*, MNRAS **355**, 1042 (2004).

28. R.I. Epstein *et al.*, Nature **263**, 198 (1976).
29. J. Linsky, Space Sci. Rev., **106**, 49 (2003).
30. B.E. Wood *et al.*, Astrophys. J. **609**, 838 (2004).
31. B.D. Fields, Astrophys. J. **456**, 678 (1996).
32. D. Kirkman *et al.*, Astrophys. J. **529**, 655 (2000).
33. Y.I. Izotov *et al.*, Astrophys. J. **527**, 757 (1999).
34. K.A. Olive and E. Skillman, Astrophys. J. **617**, 29 (2004).
35. V. Luridiana *et al.*, Astrophys. J. **592**, 846 (2003).
36. N. Christlieb *et al.*, Nature **419**, 904 (2002).
37. M. Spite and F. Spite, Nature **297**, 483 (1982).
38. S.G. Ryan *et al.*, Astrophys. J. **523**, 654 (1999).
39. E. Vangioni-Flam *et al.*, New Astron. **4**, 245 (1999).
40. S.G. Ryan *et al.*, Astrophys. J. Lett. **530**, L57 (2000).
41. P. Bonifacio *et al.*, Astron. & Astrophys. **390**, 91 (2002).
42. J. Meléndez and I. Ramirez, Astrophys. J. Lett. **615**, 33 (2004).
43. M.H. Pinsonneault *et al.*, Astrophys. J. **574**, 389 (2002).
44. D.S. Balser *et al.*, Astrophys. J. **510**, 759 (1999).
45. K.A. Olive *et al.*, Astrophys. J. **479**, 752 (1997).
46. E. Vangioni-Flam *et al.*, Astrophys. J. **585**, 611 (2003).
47. H. Reeves *et al.*, Astrophys. J. **179**, 909 (1973).
48. M. Fukugita *et al.*, Astrophys. J. **503**, 518 (1998).
49. R. Cen and J.P. Ostriker, Astrophys. J. **514**, 1 (1999).
50. G. Jungman *et al.*, Phys. Rev. **D54**, 1332 (1996).
51. M. Tegmark *et al.*, Phys. Rev. **D63**, 043007 (2001).
52. D.N. Spergel *et al.*, Astrophys. J. Supp. **148**, 175 (2003).
53. A. Blanchard *et al.*, Astron. & Astrophys. **412**, 35 (2003).
54. K. Jedamzik and J.B. Rehm, Phys. Rev. **D64**, 023510 (2001).
55. D.J. Fixsen *et al.*, Astrophys. J. **473**, 576 (1996).
56. R. Trotta and S.H. Hansen, Phys. Rev. **D69**, 023509 (2004).
57. P.C. Stancil *et al.*, Astrophys. J. **580**, 29 (2002).
58. R.H. Cyburt *et al.*, Phys. Lett. **B567**, 227 (2003).
59. G. Steigman *et al.*, Phys. Lett. **B66**, 202 (1977).
60. R. H. Cyburt *et al.*, Astropart. Phys. **23**, 313 (2005).
61. K.A. Olive and D. Thomas, Astropart. Phys. **11**, 403 (1999).
62. E. Lisi *et al.*, Phys. Rev. **D59**, 123520 (1999).
63. V. Barger *et al.*, Phys. Lett. **B566**, 8 (2003).

64. E.W. Kolb *et al.*, Phys. Rev. **D33**, 869 (1986);
    F.S. Accetta *et al.*, Phys. Lett. **B248**, 146 (1990);
    B.A. Campbell and K.A. Olive, Phys. Lett. **B345**, 429 (1995);
    K.M. Nollett and R. Lopez, Phys. Rev. **D66**, 063507 (2002);
    C. Bambi *et al.*, Phys. Rev. **D71**, 123524 (2005).
65. K.A. Olive *et al.*, Nucl. Phys. **B180**, 497 (1981).
66. J. Ellis *et al.*, Phys. Lett. **B167**, 457 (1986).
67. L.M. Krauss *et al.*, Phys. Rev. Lett. **71**, 823 (1993).
68. J.A. Morgan, Phys. Lett. **B102**, 247 (1981).
69. E.W. Kolb *et al.*, Phys. Rev. **D34**, 2197 (1986);
    J.A. Grifols and E. Massó, Mod. Phys. Lett. **A2**, 205 (1987);
    K.S. Babu *et al.*, Phys. Rev. Lett. **67**, 545 (1991).
70. A.D. Dolgov *et al.*, Nucl. Phys. **B524**, 621 (1998).
71. K. Enqvist *et al.*, Nucl. Phys. **B373**, 498 (1992);
    A.D. Dolgov, Phys. Reports **370**, 333 (2002).
72. K. Sato and M. Kobayashi, Prog. Theor. Phys. **58**, 1775 (1977);
    D.A. Dicus *et al.*, Phys. Rev. **D17**, 1529 (1978);
    R.J. Scherrer and M.S. Turner, Astrophys. J. **331**, 19 (1988).
73. D. Lindley, MNRAS **188**, 15 (1979), Astrophys. J. **294**, 1 (1985).
74. J. Ellis *et al.*, Nucl. Phys. **B259**, 175 (1985).
75. J. Ellis *et al.*, Nucl. Phys. **B373**, 399 (1992);
    R.H. Cyburt *et al.*, Phys. Rev. **D67**, 103521 (2003).
76. M.H. Reno and D. Seckel, Phys. Rev. **D37**, 3441 (1988);
    S. Dimopoulos *et al.*, Nucl. Phys. **B311**, 699 (1989);
    K. Kohri *et al.*, Phys. Rev. **D71**, 083502 (2005).
77. S. Sarkar and A.M. Cooper, Phys. Lett. **B148**, 347 (1984).
78. S. Dodelson *et al.*, Phys. Rev. **D49**, 5068 (1994).
79. J.A. Grifols *et al.*, Phys. Lett. **B400**, 124 (1997).
80. S. Weinberg, Phys. Rev. Lett. **48**, 1303 (1979).
81. M. Bolz *et al.*, Nucl. Phys. **B606**, 518 (2001).
82. G. Coughlan *et al.*, Phys. Lett. **B131**, 59 (1983).
83. N. Arkani-Hamed *et al.*, Phys. Rev. **D59**, 086004 (1999).
84. L. Randall and R. Sundrum, Phys. Rev. Lett. **83**, 3370 (1999).
85. J.M. Cline *et al.*, Phys. Rev. Lett. **83**, 4245 (1999);
    P. Binetruy *et al.*, Phys. Lett. **B477**, 285 (2000).